\documentclass{edp-conf}
\usepackage{graphicx}
\usepackage{amssymb}
\def\r){\right)}
\def\l({\left(}

\def\M{{\cal M}}
\def\MM{{\cal M}^2}
\def\rsh{r_{\rm sh}}
\def\rs{r_{*}}

\begin{document}

\TitreGlobal{SF2A 2005}

\title{The Advective-Acoustic instability in type II Supernovae}
\author{GALLETTI, P.}\address{SAp, CEA-Saclay, Orme des Merisiers 91191 Gif sur Yvette}
\author{FOGLIZZO, T.$^{1}$}
\runningtitle{The Advective-Acoustic instability in type II Supernovae}
\setcounter{page}{237}
\index{Galletti, P.}
\index{Foglizzo, T.}

\maketitle
\begin{abstract}
The puzzle of birth velocities of pulsars (pulsar kicks) could be solved by an asymmetric explosion of type II Supernovae. We propose a simple hydrodynamical mechanism in order to explain this asymmetry, through the advective-acoustic cycle (Foglizzo 2002) : during the phase of stalled shock, an instability based on the cycle between advected perturbations (entropy / vorticity) and acoustic perturbations can develop between the shock and the surface of the nascent neutron star. Eigenfrequencies are computed numerically, improving the calculation of Houck \& Chevalier (1992). The linear instability is dominated by a mode $l=1$, as observed in the numerical simulations of Blondin et al. (2003) and Scheck et al. (2004). The frequency dependence of the growth rate reveals the presence of the advective-acoustic cycle.
\end{abstract}

%

\section{Stationary accretion above a solid surface}
We consider a shocked accretion flow onto a solid surface (where the velocity of the flow is null) at a constant accretion rate. A cooling region is located above the surface and described by the generic function ${\cal L}\propto\rho^{\beta-\alpha}P^{\alpha}$ as in Houck \& Chevalier (1992), hereafter HC92. The basic equations of the flow are the continuity equation, the Euler equation and the entropy equation. This latter is : $\frac{\partial{S}}{\partial{t}}+(\vec{v}.\vec{\nabla})S+\frac{\cal L}{P}=0$, where a measure of the entropy is defined by $S\equiv1/(\gamma-1)\log(P/\rho^\gamma)$. These equations are perturbed and projected onto spherical harmonics $Y_m^l$. \\

The jump conditions at the shock $\rsh$ are given by the Rankine-Hugoniot relations for the stationary quantities. The perturbations at the shock are evaluated taking into account the displacement $\Delta \xi$ of the shock position and its velocity $\Delta v=-i\omega\Delta\xi$. In particular, the perturbations of the transverse velocity are as follows (Landau \& Lifchitz 1987) :
\begin{eqnarray}
\delta v_{\theta}&=&{v_1-v_2\over \rsh}{\partial{\Delta\xi}\over\partial{\theta}}\\
\delta v_{\phi}&=&{v_1-v_2\over \rsh \sin{\theta}}{\partial{\Delta\xi}\over\partial{\phi}}
\end{eqnarray}
where $v_1$ and $v_2$ are the pre-shock and post-shock velocities of the flow.\\
These transverse velocity perturbations at the shock are not null for non-radial perturbations. This contrasts with Eq. (51) of  HC92, who did not allow for transverse velocity perturbations at the shock.\\
 
 The eigenfrequency $\omega$ is a complex number $(\omega_r$, $\omega_i$) such that the velocity perturbation satisfies a wall type condition ($\delta v/v=0$) at the surface of the accretor. The imaginary part $\omega_i$ of the eigenfrequency is the growth rate of the perturbations.

\section{Calculations of the eigenmodes}

\begin{figure}[htb!]
\centerline{\includegraphics[width=0.5\columnwidth]{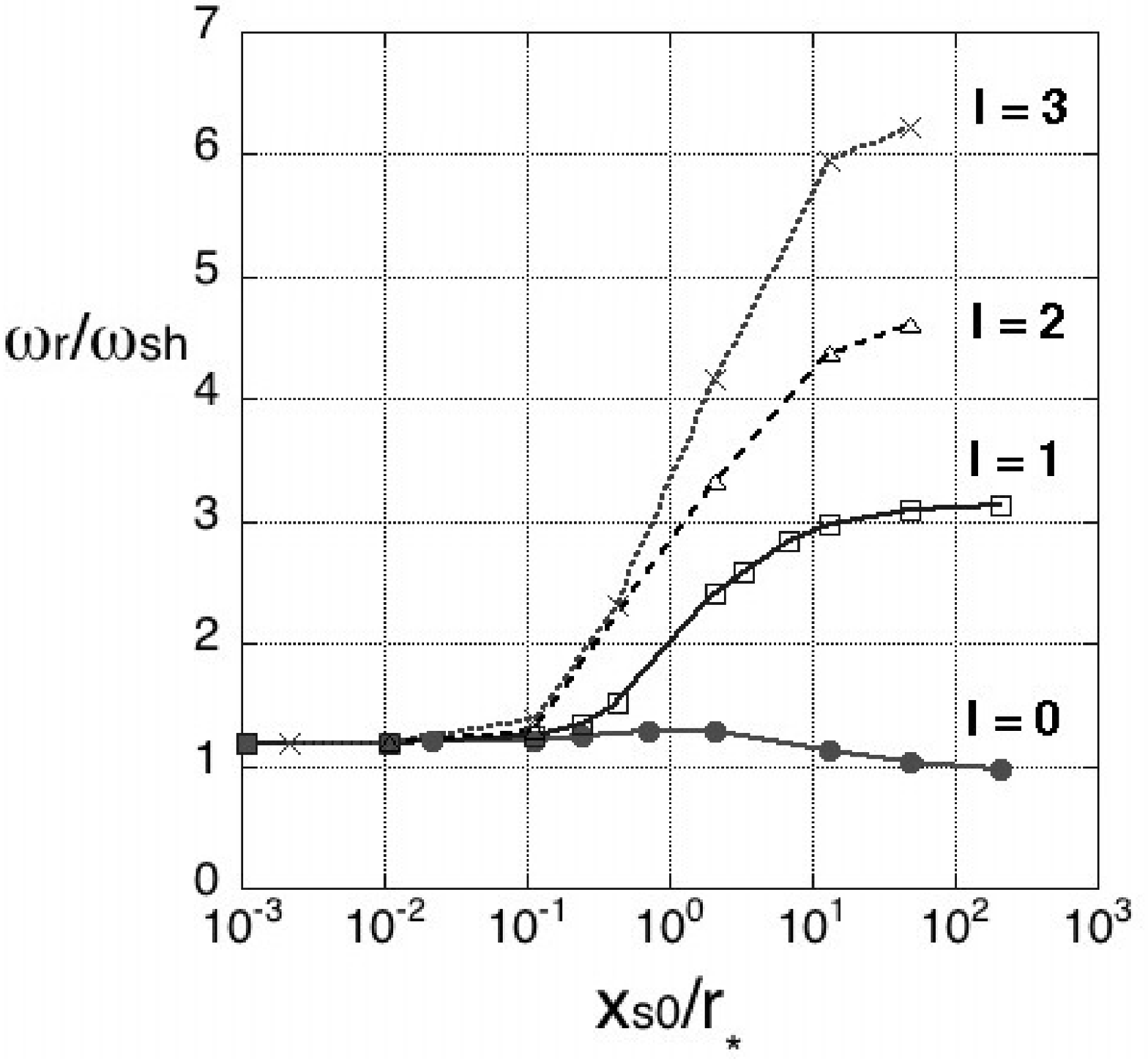}\includegraphics[width=0.5\columnwidth]{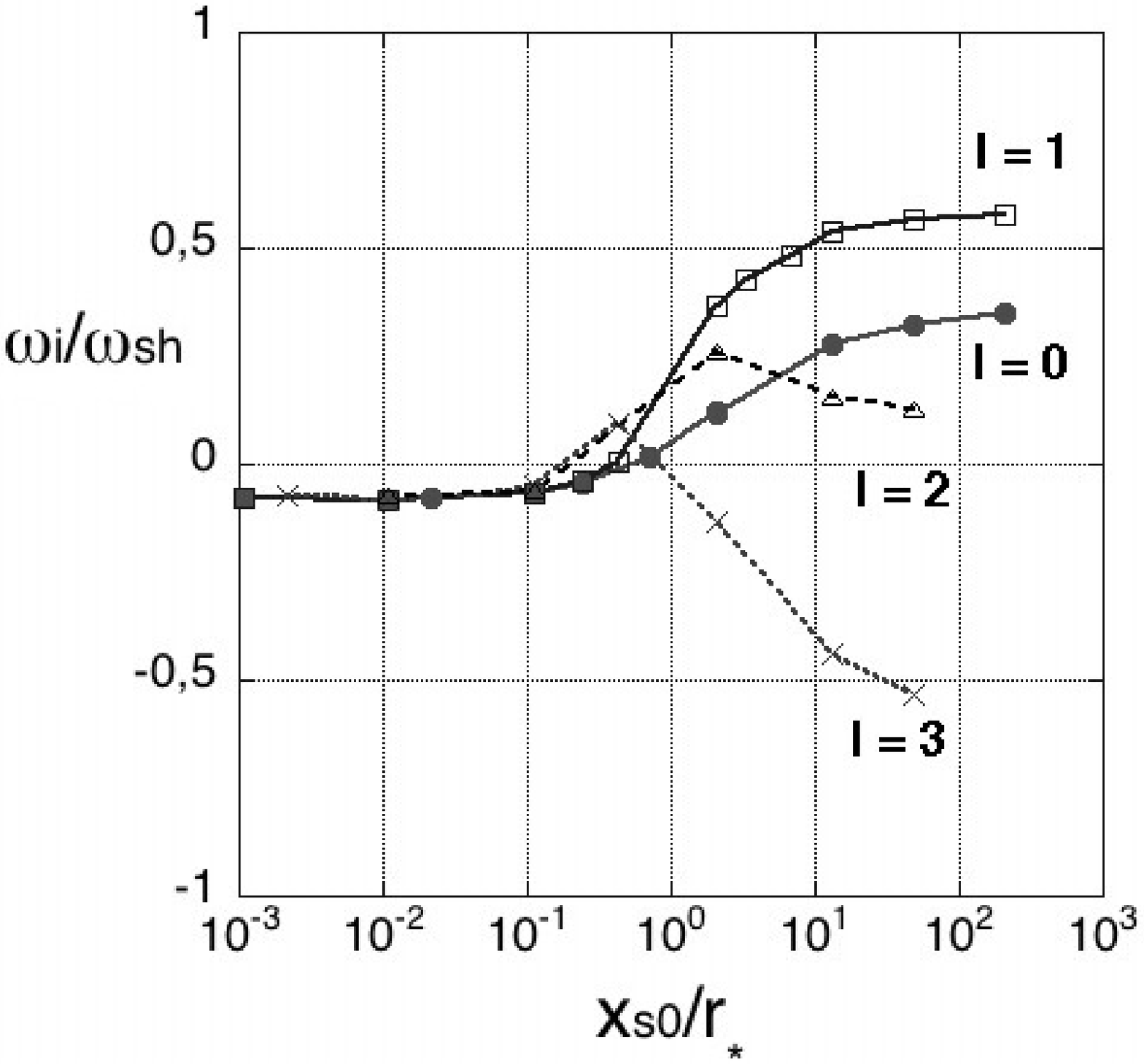}}
\caption{For $\gamma=5/3$, $\alpha=1/2$ and $\beta=2$, numerical calculations of the frequency $\omega_r$ and the growth rate $\omega_i$ in units of $\omega_{\rm sh}\equiv-v_{\rm sh}/(\rsh-\rs)$ of the fundamental modes $l=0$, 1, 2, 3 depending on the size of the cavity $x_{s0}/\rs\equiv(\rsh-\rs)/\rs$, as in Houck \& Chevalier (1992). The degree $l$ of the modes is indicated on each curve.}
\label{perturbg166l0135}
\end{figure}

We performed several calculations of the fundamental modes (an example with $\gamma=5/3$, $\alpha=1/2$ and $\beta=2$ is shown in Fig. \ref{perturbg166l0135}). The mode $l=1$ is always the most unstable. This result differs from the analysis made by HC92 because of the error in their boundary conditions at the shock.\\

The advective-acoustic instability is based on the cycle between advected perturbations (entropy / vorticity) and acoustic perturbations between the shock and the surface. A reference timescale $\tau$ for this mechanism is equal to the accretion time from the shock to the coupling region near the surface plus the time for an acoustic wave to reach the shock :
\begin {eqnarray}
\tau&\equiv&\int_{\rs}^{\rsh}{1\over1-\M}{dr\over \vert v\vert}
\end{eqnarray}
The acoustic time $t_{\rm ac}$ is defined by the time needed for an acoustic wave to propagate from the shock to the accretor and then back up to the shock, $\omega_{\rm ac}\equiv2\pi/t_{\rm ac}$ being the pulsation associated to this acoustic time :
\begin{eqnarray}
t_{\rm ac}&\equiv&\int_{\rs}^{\rsh}{2\over1-\MM}{dr\over c}
\end{eqnarray}

\begin{figure}[htb]
\centerline{\includegraphics[width=0.5\columnwidth]{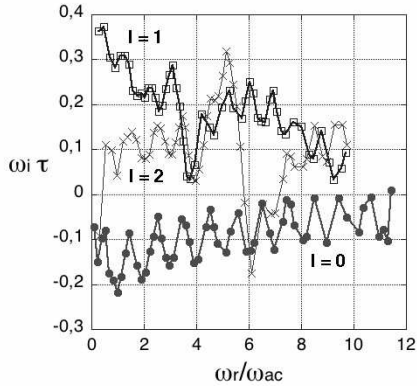}}
\caption{For $\gamma=4/3$, $\beta=2$, $\alpha=1/2$, and a cavity $\rsh/\rs\sim12$, numerical calculations of the frequency $\omega_r$ in units of $\omega_{\rm ac}$ and the growth rate $\omega_i$ in units of $\omega_{\rm sh}$ of radial $l=0$ and non-radial $l=1,2$ eigenmodes.}
\label{spectre}
\end{figure}

On Figs. \ref{perturbg166l0135}, \ref{spectre}, the growth rate $\omega_i$ is at best comparable to $\omega_{\rm sh}\sim1/\tau$ and the pulsation $\omega_r$ of the fundamental unstable modes is close to $2\pi/\tau$, as expected in the advective-acoustic mechanism. The frequency dependence of the growth rate of the eigenmodes shows an oscillatory behaviour with a period comparable to $\omega_{\rm ac}$ (Fig. \ref{spectre}). Such oscillations are expected in the advective-acoustic instability, as a consequence of the modulation of the advective-acoustic cycle by the purely acoustic cycle (Foglizzo 2002). This interpretation is confirmed by measuring the ratio $\tau/t_{\rm ac}\sim4.5-6.5$ which is comparable to the number of modes found per oscillations (Foglizzo 2002).\\
We note that for big cavities, the most unstable eigenmodes correspond to low frequencies, in the "pseudo-sound" regime ($\omega_r<\omega_{\rm ac}$). Unstable eigenmodes are also found in the acoustic regime ($\omega_r>\omega_{\rm ac}$), with a smaller growth rate however (Fig. \ref{spectre}).\\

\begin{figure}[htb]
\centerline{\includegraphics[width=0.5\columnwidth]{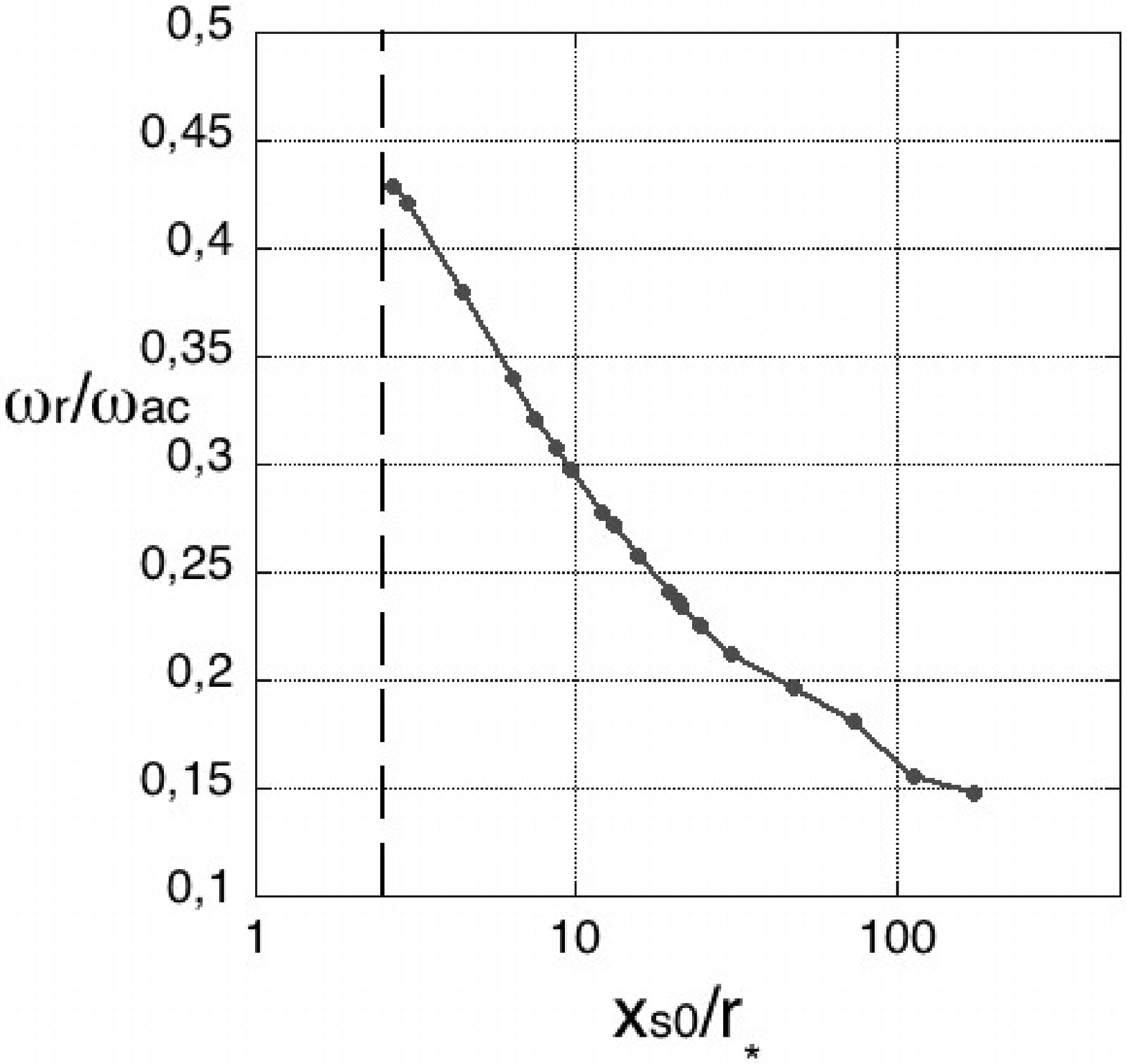}\includegraphics[width=0.5\columnwidth]{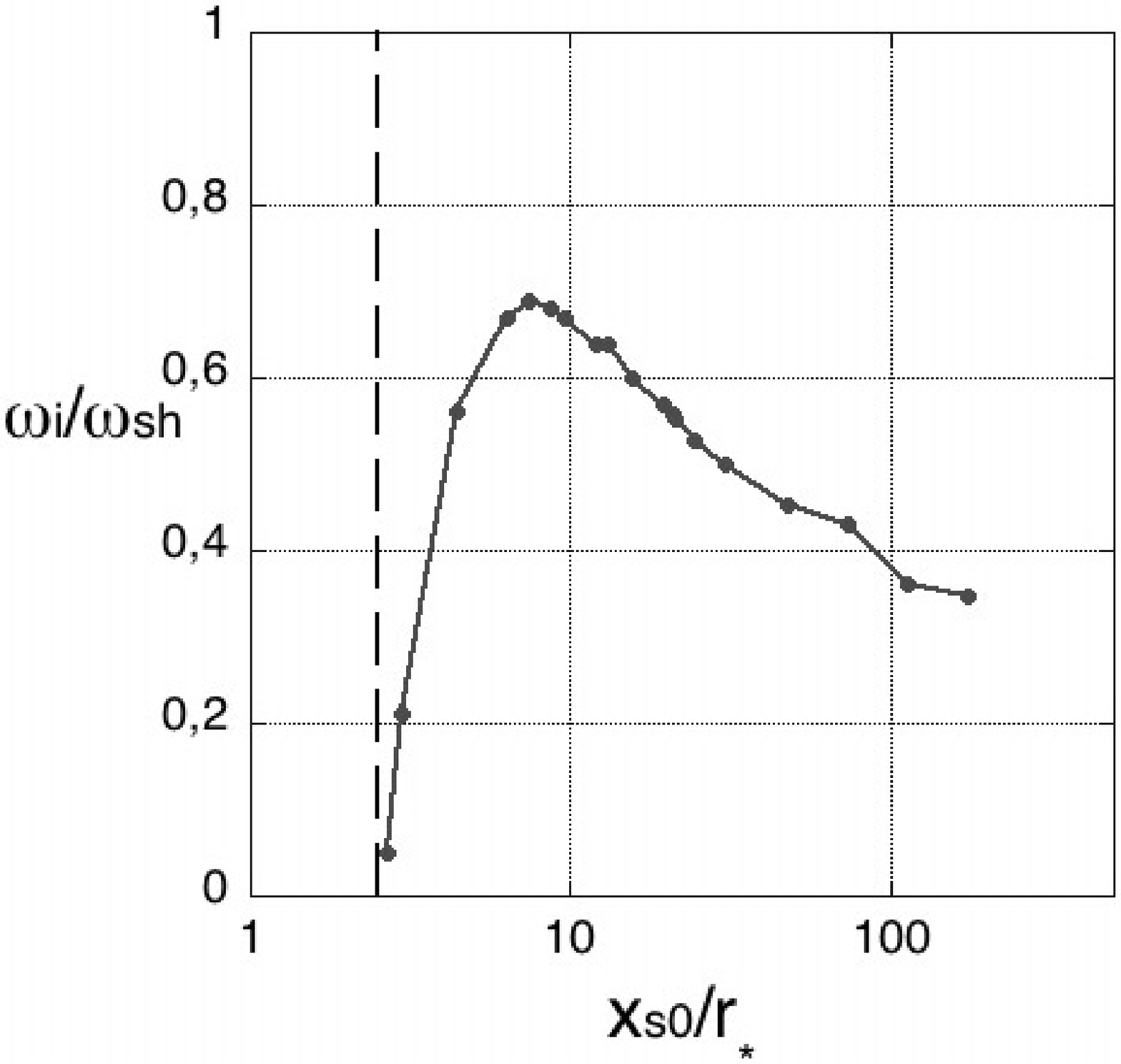}}
\caption{For  $\gamma=4/3$, $\alpha=6$ and $\beta=1$, numerical calculations of the frequency $\omega_r$ in units of $\omega_{\rm ac}$ and the growth rate $\omega_i$ in units of $\omega_{\rm sh}\equiv-v_{\rm sh}/(\rsh-\rs)$, of the first unstable mode $l=1$, depending on the size of the cavity $x_{s0}/\rs=(\rsh-\rs)/\rs$. The vertical dashed line correspond to $\rsh/\rs=3.5$ ($x_{s0}/\rs=2.5$).}
\label{perturbg133a6b1}
\end{figure}

On Fig. \ref{perturbg133a6b1}, for $\gamma=4/3$ and a cooling function described by $\alpha=6$, $\beta=1$, relevant to the phase of a stalled shock in type II Supernovae (Bethe \& Wilson 1985), an unstable mode $l=1$ can also be found for big enough cavities ($\rsh/\rs\gtrsim3.5$). A frequency dependence study, still in progress, shows unstable modes $l=1$ both in the "pseudo-sound" regime and in the acoustic regime. The radial modes ($l=0$) are always stable.\\

\section{Conclusion}

The stalled accretion shock above a neutron star is unstable, with a domination of a mode $l=1$ if the shock radius is large enough. The instability is interpreted as an advective-acoustic cycle modulated by a purely acoustic cycle. The instability is also found when the cooling function mimics neutrino cooling in type II Supernovae. The advective-acoustic cycle is thus a good candidate to seed an asymmetric explosion which could lead to an important birth velocity of the neutron star.

%
%
%
%
%
%



\end{document}